\documentclass{sf2a-conf}
\usepackage{graphicx}
\usepackage{epsfig,epsf}

\def\grs{GRS $1915$+$105$}
\def\X1550{XTE J$1550$--$564$}
\def\J1655{GRO J$1655$--$40$}

\def\integral{{\it{INTEGRAL}}}
\def\rxte{{\it{RXTE}}}
\begin{document}

\TitreGlobal{SF2A 2006}

\title{Integrally  monitoring GRS 1915+105 with simultaneous INTEGRAL, RXTE, Ryle and 
Nan\c{c}ay observations}
\author{Rodriguez, J.}\address{CEA Saclay, DSM/DAPNIA/SAp, Orme des Merisiers, F-91191 Gif sur Yvette, France}
\author{Pooley, G.}\address{Cambridge, UK}
\author{Hannikainen, D.C.}\address{Observatory, University of Helsinki, Finland}
\author{Lehto, H.J.}\address{Tuorla Observatory, Turku Finland \&  Nordita, Denmark}
\runningtitle{INTEGRAL monitoring of GRS 1915+105}
\setcounter{page}{237}
\index{Rodriguez, J.}
\index{Pooley, G.}
\index{Hannikainen, D.C.}
\index{Lehto,H.}

\maketitle
\begin{abstract}
We report here the results of 2  observations performed simultaneously  with \integral, 
\rxte, the Ryle and Nan\c{c}ay radio telescopes. These observations  belong to 
the so-called $\nu$ and $\lambda$ classes of variability during which  a high level 
of correlated X-ray and  radio variability is observed.
We study the connection between the accretion processes seen in the X-rays, and the 
ejections seen in radio. 
By observing an ejection during class $\lambda$, we generalise the fact that the discrete ejections 
in GRS 1915+105 occur after sequences of soft X-ray dips/spikes. We then identify the most 
likely trigger of the ejection through a spectral 
approach to our \integral\ data.  We show that each ejection is very probably 
the result of the ejection of a Comptonising medium responsible for the hard X-ray 
emission seen above 15 keV with \integral.
\end{abstract}
%
\section{Introduction}
GRS~1915+105 is the most active microquasar of our Galaxy. 
An extensive review on this source can be found in Fender \& Belloni (2004).
To summarize, \grs\ hosts a black hole (BH) of  14.0 $\pm$ 4.4 M$_{\odot}$ 
(Harlaftis \& Greiner 2004),
it is one of the brightest X-ray sources in the sky and it is a source of superluminal 
ejection (Mirabel \& Rodriguez 1994), with true velocity of the jets $\geq 0.9c$.
 The source is also known to show a compact jet during its periods of 
low steady levels of emission (e.g. Fuchs et al.\ 2003). 
Multi-wavelength coverages from radio to X-ray  have shown a
clear but complex association between the soft X-rays and radio/IR behaviours.
Of particular relevance is the existence  of radio QPO in the range 20--40 min  
associated with the X-ray variations on the same time scale (e.g. Mirabel et al.\ 98). These
so called ``30-minute cycles'' were interpreted as being due to small ejections of material from the
system, and were found to correlate with the disc instability, as
observed in the X-ray band.\\
\indent Extensive  observations at X-ray energies with  {\it RXTE} allowed   
Belloni et al.\ (2000) to  classify all the observations into
12 separate classes (labelled with greek letters), which could be interpreted 
as transitions between
three basic states (A-B-C): a hard state and two softer states.  These
spectral changes are, in most of the classes, interpreted as reflecting 
the rapid disappearance
of the inner portions of an accretion disc, followed by a slower
refilling of the emptied region (Belloni et al.\ 1997).  \\
\indent The link between the accretion and ejection processes is, however, far from 
being understood and different kind of models are proposed to explain all 
observational facts 
including the X-ray low (0.1-10 Hz) frequency QPOs. It should be added that 
until the launch of \integral, 
all studies had been made below 20 keV with the PCA onboard \rxte, bringing, thus, only few 
constraints to the behaviour of  the hard X-ray emitter (hereafter called corona).
 Since the launch of \integral\ in late 
2002 we have monitored  GRS 1915+105 
with long exposure ($\sim100$ ks) pointings. All the observations have been conducted simultaneously 
with other instruments, in particular \rxte\ and the Ryle Telescope, and in some cases with 
others (Spitzer, Nan\c{c}ay, GMRT, Suzaku,...), with the aim of understanding the 
physics of the accretion-ejection phenomena, including,  for the first time, the behaviour of
the source seen above 20 keV up to few hundred keV. We report here the results obtained during 
observations showing sequences of  X-ray hard dips/soft spikes (hereafter cycles), 
followed by radio flares.

\section{First occurence of a cycle during an \integral\ observation: October 2004}
\begin{figure}[h]
\begin{tabular}{ll}
\epsfig{file=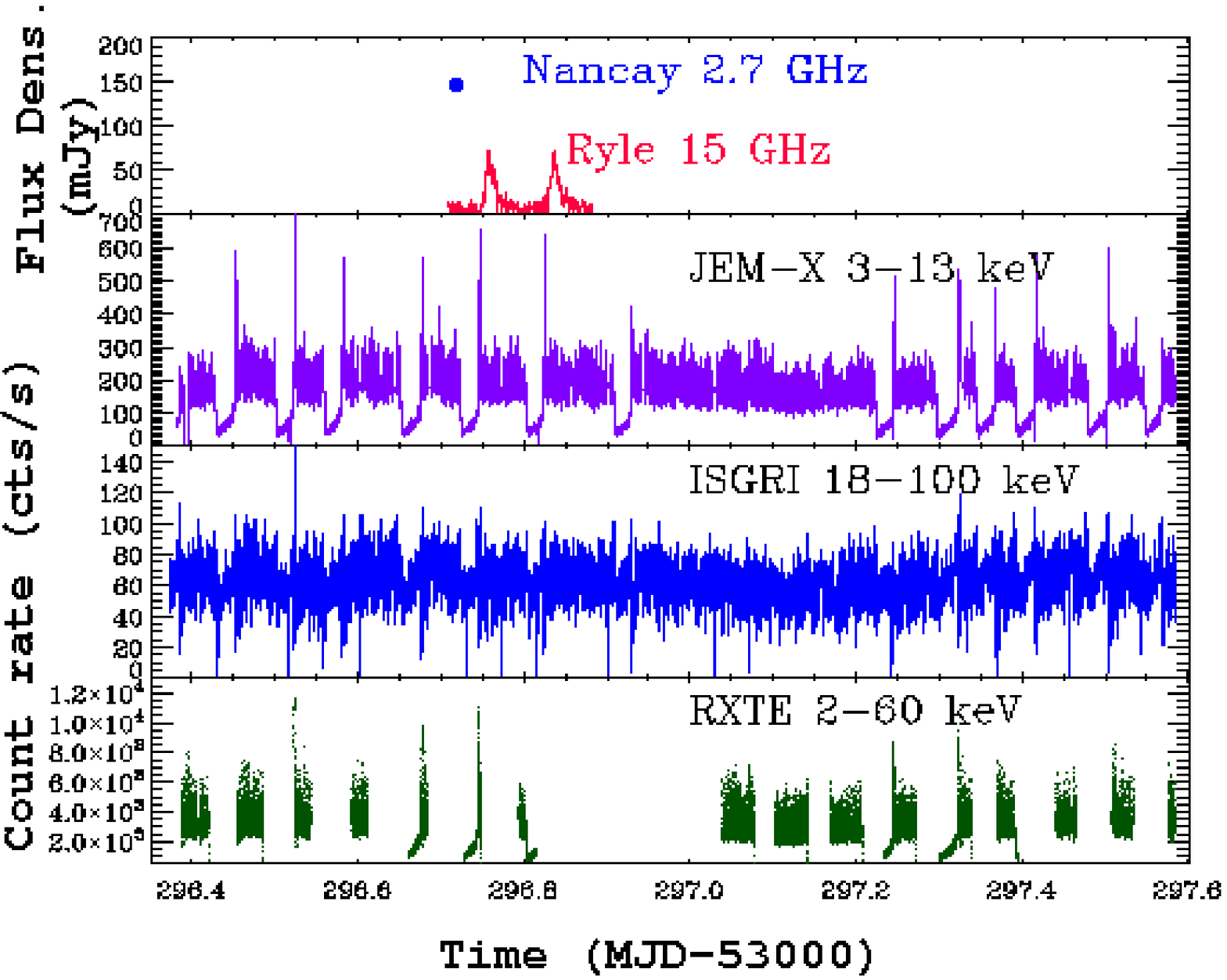,width=8cm}&
\epsfig{file=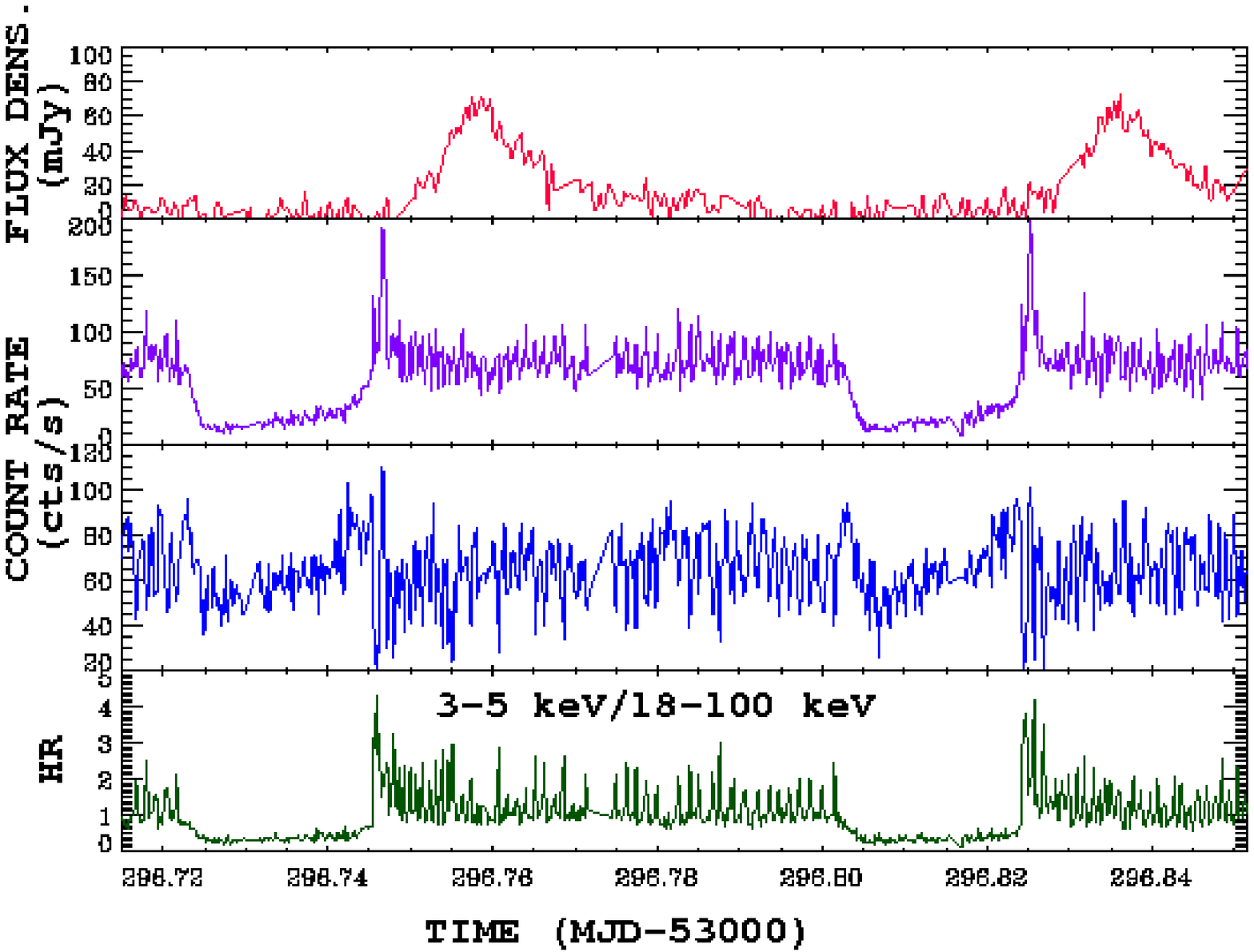,width=8cm}\\
\end{tabular}
\caption{{\bf Left:} Multiwavelength light curves of GRS 1915+105 on October 18-19 2004. 
On two occasions, 
one can see a  clear sequence X-ray dip-spike followed by a radio flare seen by 
the Ryle Telescope. {\bf Right:} Detail of the moment showing the 2 radio flares. From 
top to bottom, Ryle, JEM X 3-5 keV and ISGRI 18-100 keV light curves, the bottom 
panel represents the 3-5/18-100 keV hardness ratio.}
\label{fig:oct04}
\end{figure}

Fig. \ref{fig:oct04} shows the Nan\c{c}ay (2.7 GHz), Ryle (15 GHz), \integral\ JEM-X (3-13 keV) and
ISGRI (18-100 keV), and the \rxte\ (2-60 keV) light curves. The delay between the return to the 
high level of soft X-ray (the spike) and the peak of the radio flares is 0.28 hours and 
0.26 hours for the first and the 
second flares respectively. Zooming on the two cycles preceding  the 2 radio flares 
one can see that in each case the return to 
a high level of soft X-ray is quite complex. In all cycles, while during the dip 
both the soft and the hard X-ray seem to evolve simultaneously (as illustrated by 
the rough constancy of the hardness ratio) the major spike is preceded by a precursor 
which is associated to a very short dip (Fig. \ref{fig:oct04} left). In order 
to  understand better the spectral evolution 
during he cycle, and constrain the possible connection with  the 
ejection, we divided each sequence of dip/spike into 4 intervals from which average 
spectra were extracted and analysed. Interval A corresponds to the main dip, it has a low 
3-5/18-100 keV hardness ratio and a low soft X-ray flux. Interval B corresponds to the 
precursor spike (that is present in all cycles), interval C the dip that follows 
immediately after, and interval D corresponds to the major spike, that seems to 
precede to any interval of $\rho$-type variations, between each cycle. Note that these A, B, C, D
do not relate to the spectral states identified by Belloni et al. (2000).
We fitted the four resultant spectra with the same model consisting of an accretion disc
modeled by the {\tt{ezdisk}} model of {\tt{XSPEC}} plus a comptonised component 
({\tt{comptt}}), modified by interstellar absorption (fixed to $6\times 10^{22}$ 
cm$^{-2}$). While the details of the results are reported elsewhere (Rodriguez et al.\ 
in prep.), the best model allowed us to estimate the 3-50 keV relative unabsorbed fluxes
of each spectral component, and thus study their evolution through the cycle. We can indicate that
through the A-B-C-D sequence the accretion disc gets monotonically closer to the compact object,
with an increasing flux, while the 3-50 keV flux of the  ``corona'' is reduced by a factor 
2.4 between B and C.

\section{Ejection during a class $\lambda$ for the first time: November 2004}
\begin{figure}[h]
\begin{tabular}{ll}
\epsfig{file=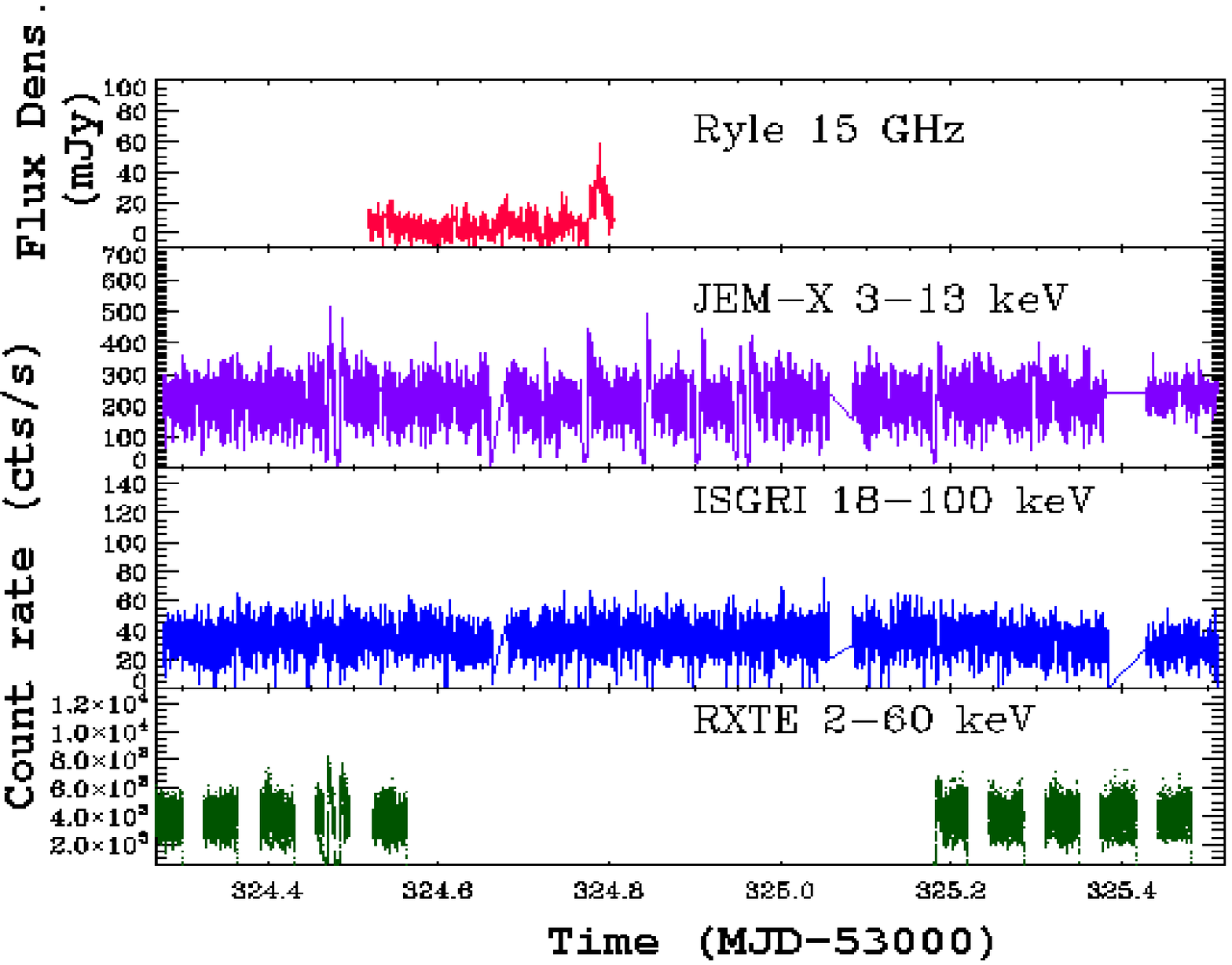,width=8cm}&
\epsfig{file=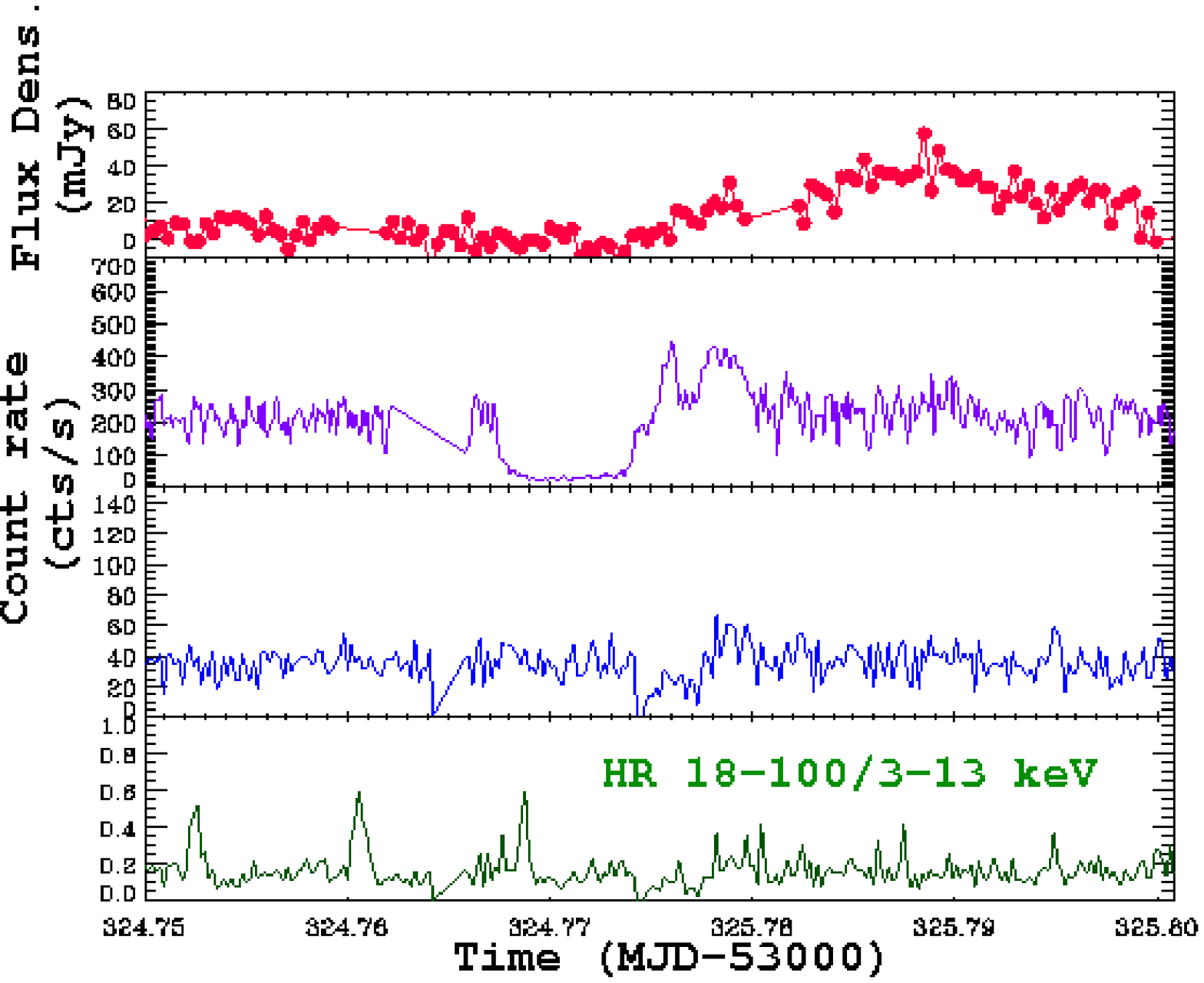,width=8cm}\\
\end{tabular}
\caption{{\bf Left:} Multiwavelength light curves of GRS 1915+105 on Nov. 15-16 2004. 
One can see a  clear sequence X-ray dip-spike followed by a radio flare seen 
by the Ryle Telescope. {\bf Right:} Detail of an X-ray cycle and associated radio flare. From 
top to bottom, Ryle, JEM X 3-5 keV and ISGRI 18-100 keV light curves, and 18-100/3-13 keV hardness ratio.}
\label{fig:Nov04}
\end{figure}

While the observation of ejections during classes showing spectrally hard dips is not very new, and 
seems to always occur after $>$100s hard dips (Klein-Wolt et al.\ 2002), no 
generalisation had ever been drawn. The reason is that this had never been observed during the 
long hard dips of class $\lambda$. During our observation on November 15-16 2004 the source is 
found in the $\kappa$ to $\lambda$ class. On one occasion, after a hard dip, a radio 
flare, which we interpret as being due to an ejection of material, is detected at 15 GHz with 
the Ryle (Fig. \ref{fig:Nov04}). The delay between the recovery of the soft X-rays seen by 
JEM-X and the radio flare is 0.31 hour. \\
\indent As for the previous observation if we zoom on the 
interesting part at X-ray energies, the cycle is in fact composed
of a hard dip, a precursor spike, a secondary dip after the spike, and the return to 
a high level of soft X-ray flux and variability (Fig. \ref{fig:Nov04}). Here we divided 
this unique sequence into 4 intervals: main dip, precursor, short dip, and major spike, 
that were fitted in 
{\tt{XSPEC}} with the same physical model as the previous observation. Again we focus on the 
evolution of the relative unabsorbed fluxes to draw our conclusions. As in the previous class, the 
disc gets closer to the compact object throughout the cycle with an
increasing flux, while the 3-50 keV flux of the corona is reduced by at least a factor 2.7, and up to 
a factor 10 if all spectral parameters are left free to vary in the fits.

\section{Discussion}
Altough the 2 observations show clear differences, visible in the shape of the 
light curves, the evolution of the hardness 
ratio (Fig. \ref{fig:oct04} \& \ref{fig:Nov04}), that of the spectral parameters, 
 the similarity of the cycles in their sequences,  indicate that the physics connecting
accretion and ejection is the same.
In class $\nu$ the disc seems to approach the compact object and gets brigther, 
throughout the A-B-C-D sequence, it does the same in class $\lambda$. In both cases
the evolution seems to be marked by a real change in the physical properties of the 
Comptonizing plasma (corona) which result in  a change of the spectral state. More importantly, 
when studying the evolution of the relative fluxes of
the 2 components in the various intervals, while in all cases the 3-50 keV disc flux increases
from the dip to the main spike, the Compton flux decreases by a factor of $\sim2.4$ in 
class $\nu$, and higher than 2.7 in class $\lambda$. This behaviour seems to suggest that 
the dip following the precursor spike in the 2 classes, is due to the disappearance of 
the so-called ``corona'', and not simply to a pivoting of the spectrum. \\
\indent Given the disappearance of one emitting medium on the one hand, and the appearance of ejected
material on the other hand, it is tempting to consider that the ejection
occurs at the precursor spike, and is therefore the result of the ejection of the corona. In class
$\beta$ it has been shown that the X-ray spike half way through the dip is the trigger of the 
ejection (Mirabel et al.\ 1998). In addition, Chaty (1998)  has further proposed that during class 
$\beta$, the spike corresponds to the disappearance of the corona, suggesting that the coronal 
material is ejected in this particular class too. Rodriguez et al.\ (2002) reached the same conclusions
 comparing the behaviour of the source at low and hard X-ray energies during a class 
$\alpha$ observation.  The case of GRS 1915+105, although spectacular, is not unique. 
Rodriguez et al.\ (2003) have come to the conclusion that in the microquasar  
XTE J1550$-$564 during its 2000 outburst the corona is ejected at the peak of the outburst and 
further detected in radio. The main difference with 
\grs\ are the constants of time over which the ejections occur. While in XTE J1550$-$564, 
one ejection only is seen after the peak of the outburst at soft X-ray energies, 
in GRS 1915+105 those kind of events can occur several times within few hours 
(Fig. \ref{fig:oct04}). The general behaviour, however, seems to be the same. The ejection 
is coincident with a disappearance of one of the emitting medium, leading to the  
conclusion that it is this media, the corona, that is blown away and detected later at 
radio wavelengths.\\
\indent In addition to confirming this proposition by including a precise analysis of the hard 
X-rays, our observations allow us to generalise the connection of the accretion-ejection 
phenomena in GRS 1915+105 by observing 
an ejection during a class $\lambda$ observation for the first time. We also provide evidence
that in at least two cases showing dip-spike sequences, the return to a high level of soft X-ray,
or the major spike, has a complex structure, and can be subdivided into a precursor spike 
followed by a rapid dip and a major flare. Our analysis suggests that the ejection is triggered 
at  the precursor spike, similarly to class $\beta$ in which the ejection is triggered at the 
soft spike half way 
through the soft X-ray dip (Mirabel et al.\ 1998). Preliminary analysis of an \integral\ observation 
during which ejections are observed show that the time delay between the 
soft X-ray spike and the peak of the radio flare at 15 GHz is of the same order of magnitude  
in class $\beta$ (about 0.3 hours Rodriguez et al.\ in prep.), vs. $\sim$ 0.27 and 0.31 hours 
in classes $\nu$ 
and $\lambda$ respectively. The relative constancy of this delay further strenghtens the 
association of the spike and the radio flare in all classes.\\
\indent Although the mechanisms giving rise to the ejections is far from being understood,
our analysis answers the important question regarding  the moment of the ejection which,
in turns,  brings interesting 
constraints on the existing models. The sequences of X-ray dip/spike could be due 
to a ``magnetic flood'' (Tagger et al.\ 2004) during which a magnetic instability develops 
during the spectrally hard dips, advects magnetic flux in the inner regions of the disc. 
The X-ray spike 
and associated ejection would be due to a reconnection event allowing the system to get rid 
of the accumulated magnetic flux, blowing away the corona, and leading to the evolution to 
a soft X-ray state. Many points remain unsolved, though. In particular, our analysis poses the 
question on the relation between the approach of the disc and the ejection of the 
corona. Is one consequence of the other?
\indent Our multiwavelength monitoring will continue in the future with the main goal to obtain 
more sequences of dip-spike radio flare sequences which will therefore allow us to have better 
statistics on the occurence of radio flares and their connection to the X-rays, and hard X-rays 
behaviour, and hopefully better constrain  the models.

\section*{Acknowledgements}JR warmly thanks Jerome Chenevez and Carol Anne Oxborrow for their 
useful help with the JEM-X data reduction. JR also acknowledges  useful discussions with 
Marion Cadolle Bel, St\'ephane Corbel and Michel Tagger.

\end{document}